\documentclass[%
 apl,
 amsmath,amssymb,
reprint,%
floatfix, %  should fix the warning "A float is stuck (cannot be placed); try class option [floatfix]" in most cases
]{revtex4-2}

\usepackage{graphicx}% Include figure files
\usepackage[caption=false]{subfig} %subfigure files

\usepackage{dcolumn}% Align table columns on decimal point
\usepackage{bm}% bold math
\usepackage[utf8]{inputenc}
\usepackage[T1]{fontenc}

\usepackage{siunitx}
\usepackage[dvipsnames]{xcolor}
\usepackage[colorlinks=true, linkcolor = blue, urlcolor = RoyalBlue, citecolor = magenta]{hyperref}
\usepackage{soul} % strike text

\usepackage{natbib}

\begin{document}

\preprint{AIP/Cond Mat}

\title[Controlled generation and detection of a thermal bias in Corbino devices under the quantum Hall regime. ]{Controlled generation and detection of a thermal bias in Corbino devices under the quantum Hall regime. }

\author{M.A. Real}
 \email{mreal@inti.gob.ar}
 \affiliation{Quantum Metrology Department, INTI, 1650 Buenos Aires, Argentina.}
 \affiliation{INCALIN, UNSAM, Argentina.}

\author{A. Tonina}%
 \affiliation{Quantum Metrology Department, INTI, 1650 Buenos Aires, Argentina.}
 \affiliation{INCALIN, UNSAM, Argentina.}%

\author{L. Arrachea}
\affiliation{Escuela de Ciencia y Tecnología and ICIFI, UNSAM, Campus Miguelete, 1650 Buenos Aires, Argentina}%
\affiliation{CONICET, Argentina.}%

\author{P. Giudici}%
 \affiliation{Instituto de Nanociencia y Nanotecnolog\'{\i}ga, CNEA-CONICET, 1650 Buenos Aires, Argentina.}%

\author{C. Reichl}
\affiliation{Solid State Physics Laboratory, ETH Z\"urich, CH-8093 Z\"urich, Switzerland.} 

\author{W. Wegscheider}
\affiliation{Solid State Physics Laboratory, ETH Z\"urich, CH-8093 Z\"urich, Switzerland.}

\author{W. Dietsche}
\affiliation{Solid State Physics Laboratory, ETH Z\"urich, CH-8093 Z\"urich, Switzerland.} 
\affiliation{Max-Planck-Institut f\"ur Festk\"orperforschung, Heisenbergstrasse 1, D-70569 Stuttgart, Germany.}

\date{\today}% It is always \today, today,

\begin{abstract}
We present an experimental technique to generate and measure a temperature bias in the quantum Hall effect of GaAs/AlGaAs Corbino samples.
The bias is generated by injecting an electrical current at a central resistive heater and the resulting radial temperature drop is determined by conductance measurements at internal and external concentric rings. 
The experimental results agree
with the predictions of numerical simulations of the heat flow through the substrate. We also compare these results with previous predictions based on the thermoelectric response of these devices.
\end{abstract}

\maketitle

% To be published in AIP or APL (https://aip.scitation.org)

%%%%%%%%%%%%%%%%%%%%%%%%%%%%%%%%%%%%%%%%%%%%%%%%% 
% Intro
%%%%%%%%%%%%%%%%%%%%%%%%%%%%%%%%%%%%%%%%%%%%%%%%%

The quantum Hall effect (QHE) \cite{klitzing1980new} has been a cornerstone to many fundamental phenomena in physics since its discovery more than 40 years ago \cite{klitzing1980new}. It also has  a key role in electrical metrology and in the 2019 redefinition of the International System of Units \cite{CGPM26}. Two very interesting sides of this effect, which motivated several insightful experiments, are its thermal \cite{venkatachalam2012local,altimiras2012chargeless,iftikhar2016primary,gurman2012extracting,bradley2016nanoelectronic,granger2009observation,melcer2022absent,jezouin2013quantum}
and thermoelectric \cite{chickering2010thermopower,chickering2013thermoelectric,VanHouten1992,Zalinge2003,endo2019spatial,Liu2018, kobayakawa2013diffusion,real2020thermoelectricity}
properties.
The related research activity, along with the study of these properties in other mesoscopic devices is  accompanied by the  development of  reliable techniques for temperature sensing \cite{Giazotto2006Mar}.

A large body of work on thermal and thermoelectric transport in the quantum Hall regime
focuses on the  edge states, where the transport takes place through a few ballistic channels.
This  is an ideal playground to study many fundamental properties from the theoretical \cite{Kane1997thermalTransport,Sanchez2015,vannucci2015interference,Roura-Bas2018Feb} and  experimental point of view. 
Prominent experimental examples are the investigation of the
chiral propagation of heat \cite{granger2009observation,venkatachalam2012local,nam2013thermoelectric}, the study of the peculiar
transport of charge and energy in some
fractional filling factors \cite{altimiras2012chargeless,banerjee2017observed},
and the detection of the quantum of thermal conductance \cite{jezouin2013quantum}. 
The main experimental strategy to induce thermal transport in  this context is the implementation of ohmic contacts at the edge states. 
Local temperature sensing relies on recording the conductance in
tunnel-coupled quantum dots in the Coulomb blockade regime, playing the role of thermometers \cite{venkatachalam2012local,altimiras2012chargeless,iftikhar2016primary,gurman2012extracting,bradley2016nanoelectronic}, or noise analysis \cite{melcer2022absent,jezouin2013quantum}.

A different scenario, albeit equally interesting, is expected to take place in the thermal  and thermoelectric transport through the Landau levels, the bulk states of the quantum Hall regime. Relatively less experimental results have been reported on these phenomena in comparison to those  on the edge states.
Efforts in this direction by Chickering et al. \cite{chickering2010thermopower}, Liu et al. \cite{Liu2018} and Endo et al. \cite{endo2019spatial} focus on implementations using the Hall bar setup. In this geometry, the thermal and thermoelectric response functions have a tensorial structure with coupled longitudinal and transverse components akin to the electrical ones.  
As highlighted in the theoretical work of Ref. \cite{barlas2012thermopower}, the electrical and thermal properties of the Landau levels can be more directly assessed in the Corbino (ring) geometry.
The reason is that an electrical and/or thermal bias radially applied in the Corbino configuration generate radial fluxes in the bulk states, which are decoupled from the transverse ones.
This motivated recent studies of thermoelectric transport
\cite{kobayakawa2013diffusion,real2020thermoelectricity,mateos2021thermoelectric,li2022hydrodynamic}, complementing earlier experimental \cite{Zalinge2003} and theoretical results \cite{giazotto2007landau}.

The goal of this contribution is the proposal of an experimental method to study the thermal and thermoelectric transport through the Landau levels of a two-dimensional electron system (2DES) under the quantum Hall regime of a Corbino device.
This consists in the controlled generation of a temperature bias in the 2DES as well as its corresponding sensing.
Importantly, because Landau levels extend throughout the bulk of the sample, edge-state techniques such as those in Refs. 
\cite{granger2009observation,venkatachalam2012local,altimiras2012chargeless,iftikhar2016primary,gurman2012extracting,bradley2016nanoelectronic}, cannot be used.

\begin{figure}
    \centering
    \includegraphics[width=\columnwidth]{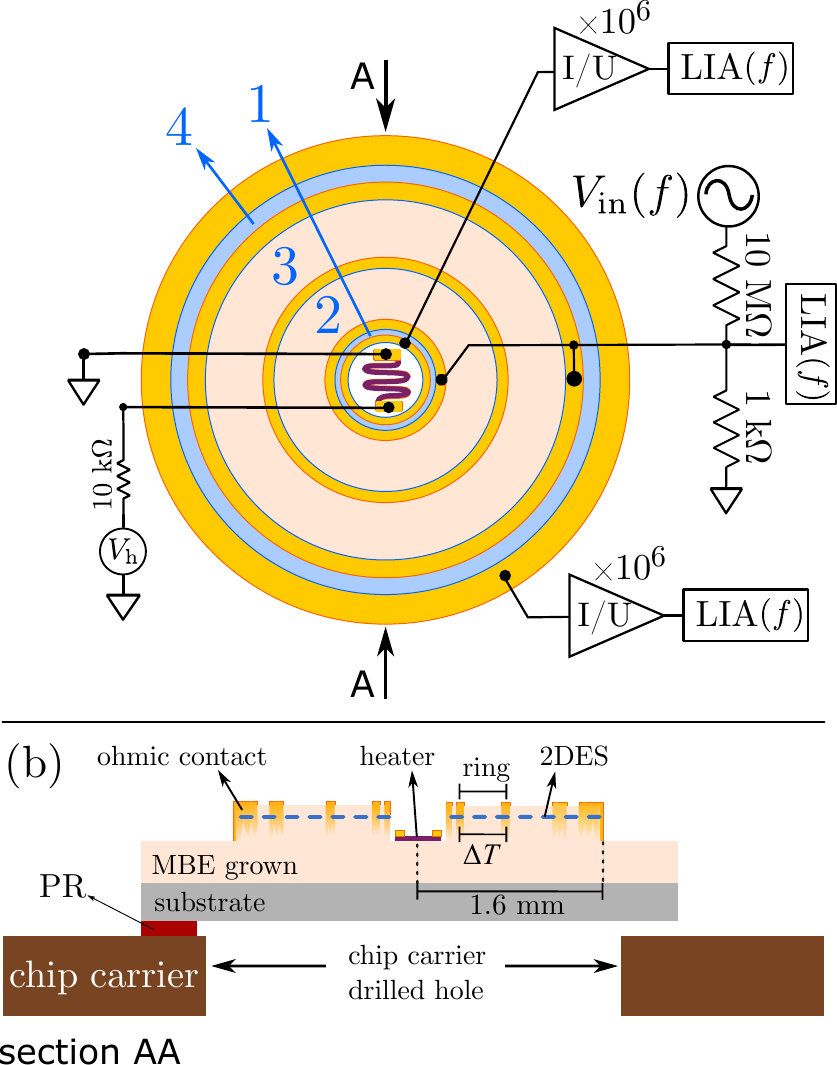}
    \caption{\textbf{Experimental setup.}
    (a) Sketch of the device. Five ohmic contacts (yellow) confine  four enumerated Corbino rings 
    (1 and 4 in light blue, 2 and 3 in light orange). 
    The voltage $V_{\textrm{h}}$ is applied to the central resistive heater to induce a temperature gradient on the substrate. The conductance of the 2DES can be measured independently in any of the 
    rings introducing a bias at the corresponding ohmic contact through a voltage divider applying a voltage $V_{\textrm{in}}(f)$, while the resulting current is measured employing two current-to-voltage linear converters. LIA denotes lock-in amplifier. 
    We focus on the Corbino rings 1 and 4, which are highlighted in light blue.
    (b) Cross-section AA, the mesa and ohmic contacts are shown. Note that the heater element is outside the mesa, we rely on the crystal to induce the temperature gradient that will impact the 2DES. 
    A hole is drilled in the chip carrier to reduce thermal contact with the sample in the Corbino area. The sample is fixed to the carrier with photo-resist (PR) only to one edge of the sample.}
    \label{fig:fig1-expSetup}
\end{figure}

We focus on the experimental setup shown in Fig.~\ref{fig:fig1-expSetup} (a), where the central region of the sample is heated using an electrical resistor and the outer rim is connected to the cold finger of a cryostat.
To generate the thermal bias, it is important to notice that the 2DES has a thermal conductivity which is much lower than that of the phonons in the substrate \cite{chickering2010thermopower,chickering2013thermoelectric}.
Hence, the latter defines the limit to establish the temperature gradient in the substrate, while the 2DES locally thermalizes with it. 
To sense the temperature drop at the 2DES, a series of five concentric electrical contacts are implemented within the Corbino disk [yellow areas in Fig. \ref{fig:fig1-expSetup} (a)]. 
These contacts define four regions of the 2DES, labeled as rings 1,2,3 and 4. We measure the conductance of rings 1 and 4 and use its dependence on temperature to sense the temperature drop between the ring 1 and 4.
We compare these measurements with numerical estimates of the temperature gradient in the substrate, based on calculations of the phononic heat flux through the substrate and find excellent agreement. 
Furthermore, we demonstrate that there exists a linear relationship between the power applied to the heater and the temperature gradient in the sample.

%%%%%%%%%%%%%%%%%%%%%%%%%%%%%%%%%%%%%%%%%%%%%%%%%%%%%%%%%%%%%%%%%%%%%%%%%%%%%%%%
%\section{Experimental setup}
%%%%%%%%%%%%%%%%%%%%%%%%%%%%%%%%%%%%%%%%%%%%%%%%%%%%%%%%%%%%%%%%%%%%%%%%%%%%%%%%

The single 2DES forms in a GaAs/AlGaAs heterostructure, grown by molecular beam epitaxy on a GaAs wafer.
Control samples measured independently at \SI{1.3}{\kelvin} in the dark, resulted in 2DES mobilities and densities of the order of \SI{2.0e7}{\cm \squared \per\volt\per\second} and \SI{3.1e-11}{\cm^{-2}}, respectively.
Using micro-processing techniques, a ring-shaped mesa and ohmic contacts of Ni-Ge-Au are implemented on square-cleaved samples. 
The final configuration has four concentric rings defined by the five ohmic contacts. 
The internal and external diameters of the full Corbino structure hosting the four rings are, respectively,  \SI{0.4}{\milli\meter} and \SI{3.2}{\milli\meter}.
The central AuPd heater is placed outside the mesa. The sample is placed on a chip carrier, which is in contact with a $^3$He cold finger cryostat of \SI{250}{\milli\kelvin} base temperature.
The chip carrier is drilled to contain a central hole with a diameter of \SI{3.5}{\milli\meter} in order to minimize the thermal contact with the substrate. Only one side of the sample is glued to it [see PR in the left hand side of Fig.\ref{fig:fig1-expSetup} (b)].
The temperature at the cold finger is recorded with a Cernox sensor.
The magnetic field is generated by a \SI{14}{\tesla} superconducting magnet.

Magneto-transport measurements can be performed independently in any of the 4 rings by suitably connecting the two consecutive ohmic contacts that define the ring to be measured.
We have confirmed that the four rings exhibit independent responses.
For the measurements of the conductance ($G$) a \SI{113}{Hz} ac voltage $V_{\text{in}}(f)$ is applied to the system via a voltage divider. 
The induced electrical current is then measured by a lock-in amplifier (LIA) and a current-to-voltage amplifier (IUamp), whose frequency response has been previously verified to be flat within the measurement range.
We focus on measurements at a fixed magnetic field \SI{1.6}{\tesla}, corresponding to Landau level $N = 4$, because the conductance at this magnetic field shows a clear dependence on the temperature, allowing a good resolution for the temperature calibration. We will further discuss this choice below.

%%%%%%%%%%%%%%%%%%%%%%%%%%%%%%%%%%%%%%%%%%%%%%%%%%%%%%%%%%%%%%%%%%%%%%%%%%%%%%%%
%\section{Results and discussion}
%%%%%%%%%%%%%%%%%%%%%%%%%%%%%%%%%%%%%%%%%%%%%%%%%%%%%%%%%%%%%%%%%%%%%%%%%%%%%%%%
\begin{figure}[htp]

    \subfloat{%
      \includegraphics[width=\columnwidth]{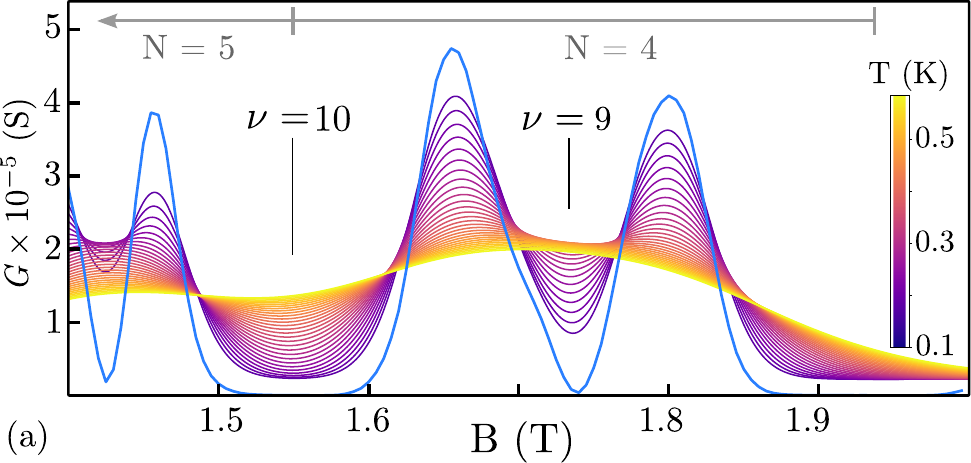}%
    }

    \subfloat{%
      \includegraphics[width=\columnwidth]{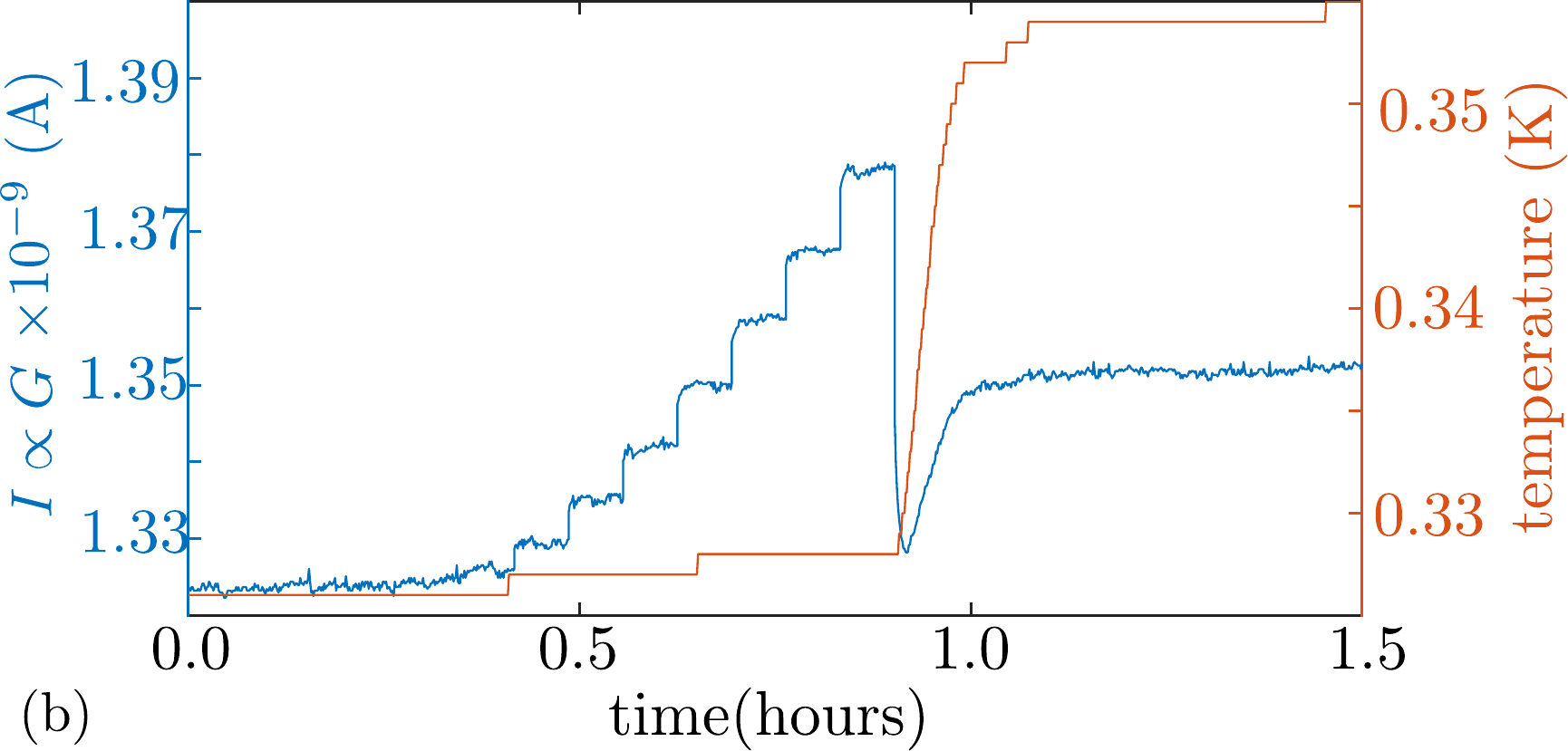}%
    }

    \caption{ (a) The measured conductance of Corbino ring 1 as a function of the magnetic field at a base temperature of \SI{260}{mK} is shown in blue.
    The other curves are calculated using Eq.~\ref{eq:G_T} with a transmission function inferred from the experimental data, in a range of temperatures from \SIrange{0.1}{0.6}{\kelvin}. 
    The filling factors $\nu$ and Landau levels N are indicated.
    (b) An example of a raw measurement is shown, where the current $I$ (proportional to the conductance $G$) and base 
    temperature measured on the cryostat cold finger Cernox sensor
    are displayed against time $t$. 
    At $t=0$ a temperature $T$ is set with the heater off. The value of $T$ is obtained from the sensor integrated in the cold finger. The different steps for $0<t<\SI{0.9}{\hour}$ correspond to changes in the heater power. At $t= \SI{54}{\minute}$ a new temperature setting is implemented and the procedure is repeated. The lag between the temperature and conductance changes is due to the thermalization time of the 2DES.}

    \label{fig:responseHeaterTemp}
    
\end{figure}

The crucial property to sense the temperature is the dependence of the magneto-conductance of the different rings as a function of the temperature. Given the transmission function $\mathcal{T}(\varepsilon)$, characterizing the transparency of the device and determined by its microscopic details, the conductance reads 
\begin{equation}
    G(T)=-\frac{e^2}{h}\int d\varepsilon \mathcal{T}(\varepsilon)  \frac{d f(T,\varepsilon)}{d \varepsilon},
\label{eq:G_T}
\end{equation}
where $f(T,\varepsilon)=\left[1+ e^{(\varepsilon-\mu)//k T)}\right]^{-1}$ is the Fermi-Dirac distribution function, which depends on the temperature $T$ and the chemical potential $\mu$. 
Our procedure is based on performing independent measurements of the conductance of the rings 1 and 4. Ideally, by using the temperature dependence predicted by Eq.~(\ref{eq:G_T}) we could define the average temperature of the 2DES in the area enclosed by each of these rings. However,
it is not possible to have an exact and accurate expression for the function $\mathcal{T}(\varepsilon)$ for each ring and for all values of $B$. 
Therefore, we introduce a fully experimental protocol to define the average temperature of a given ring on the  basis of the measurements of its conductance. One important aspect  is  to identify a regime where there is a one-to-one correspondence between $G$ and $T$. To this end, before implementing the experimental protocol, we carry out some preliminary theoretical estimates by inferring the transmission function following Ref.~\onlinecite{real2020thermoelectricity} (this approach leads to accurate results at high magnetic fields as $T \rightarrow 0$).
We show in Fig.~\ref{fig:responseHeaterTemp}~(a) results of the evolution of the conductance of ring 1 for different temperatures calculated using a transmission function inferred from the data of $G(B, T=\SI{260}{mK})$. While we are not able to accurately fit the data for any $T, B$, these plots are useful for seeing semi-quantitative features of the evolution of the conductance with $T$.
In particular, we notice that $G(B=\SI{1.6}{\tesla},T)$ is a monotonous increasing function of temperature, which is a useful property to define a one-to-one
correspondence between $G$ and $T$. For this reason, we focus on this value of the magnetic field in our analysis.

We now turn to discuss the experimental protocol. The first step is to record the conductance $G_0(T)$ without any power applied to the central heater. 
We consider bath temperatures between \SI{267}{\milli\kelvin} and  \SI{600}{\milli\kelvin}, as measured with the cold finger thermometer. 
In this way, we obtain a one-to-one relation between temperature of the 2DES and conductance. Therefore, $G_0$ becomes our calibration curve.
The second step is to record the conductance at Corbino rings 1 and 4 for different values of the heater power and bath temperatures.
After each bath temperature change, we wait at least ten minutes for thermalization, until the temperature fluctuation is lower than $\SI{1}{mK}$. When the conductance reaches a stable value we perform a set of measurements changing the DC heater power.
This process is repeated for different bath temperatures.
As an example, Fig.~\ref{fig:responseHeaterTemp} shows a set of simultaneous measurements of the conductance and temperature over time.

\begin{figure}
    \centering
    \includegraphics[width=\columnwidth]{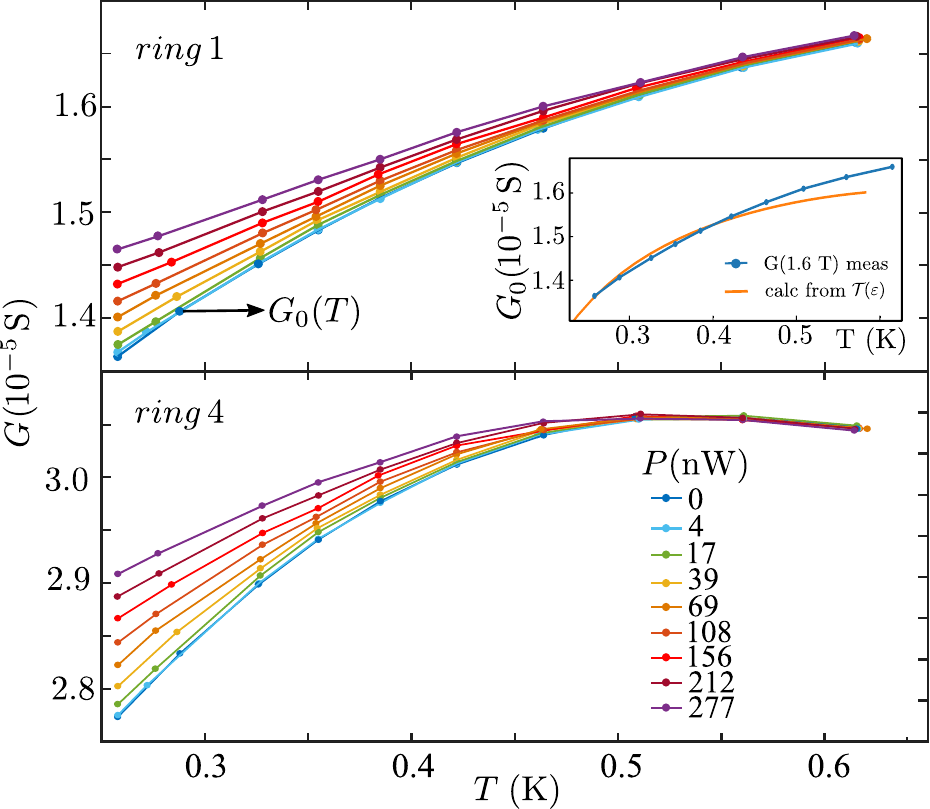}
    \caption{Conductances of ring 1 (upper panel) and ring 4 (lower panel) for different bath temperatures $T$ are shown. Each set (color) corresponds to a different DC power $P$ applied to the heater. The lines between points are interpolations. Given a power and temperature at ring $j\; (j=1,4)$ there is a corresponding zero-power (heater-off) conductance $G_0(T)$, from which the temperature change can be inferred. 
    \textit{Inset:} Calculated conductance as a function of different base temperatures for $B=\SI{1.6}{\tesla}$, as seen in Fig.~\ref{fig:responseHeaterTemp}(a) and eq.~(\ref{eq:G_T}).}
    \label{fig:calT}
\end{figure}
 
The sets of conductance measurements in the two rings at $B=\SI{1.6}{\tesla}$, for different base temperatures and for several heater powers are plotted in Fig.~\ref{fig:calT}. 
For the case of ring 1, we also show in the inset of panel (a), the measurements at zero power along with the prediction based on Eq.  (\ref{eq:G_T}), and we observe a good semi-quantitative agreement. 
The deviation should be traced back to the fact that the function ${\cal T}(\varepsilon)$ has not been inferred from data at low enough temperature.
The results for ring 4 are shown in panel (b). The  behavior of the conductance of the two rings as a function of the temperature is not exactly the same. In particular, a non-monotonous behavior observed in the ring 4 in contrast with that of ring 1. This is a consequence of inhomogeneities and charge density fluctuations along the sample. Nevertheless, we  see that below $T = \SI{0.5}{\kelvin}$ the conductance is a monotonous function of the  temperature  for all power levels in both rings. This defines a useful range of temperatures to identify a one-to-one correspondence between the conductance value and the temperature.
Here, we use the calibration of $G_0(T)$ obtained at zero heater power to define the mean local temperatures $T_{\text{ring} \, j}$ of rings $j=1,4$ at different powers since we can associate each measured conductance to a 2DES temperature.
The procedure simply corresponds to drawing a horizontal line in the corresponding panel of Fig.~\ref{fig:calT}, starting at the measured conductance of the ring  until intersecting $G_0(T)$.
The horizontal coordinate of the intersection point defines $T_{\text{ring} \, j}$.

The temperature difference $(T_{\text{ring} \, 1}- T_{\text{ring} \, 4})^{\rm meas}$ is determined from the 
estimates of the temperatures of the two rings. 
This temperature difference is obtained at fixed values of the heater power by averaging the results over a cold finger temperature range from \SIrange{260}{550}{\milli\kelvin}.  Results are shown in  Fig.~\ref{fig:deltaTpower}, where we verify a linear dependence between these two quantities. The error bars indicate the statistical errors associated with the averaging procedure.

\begin{figure}
    \centering
    \includegraphics[width=\columnwidth]{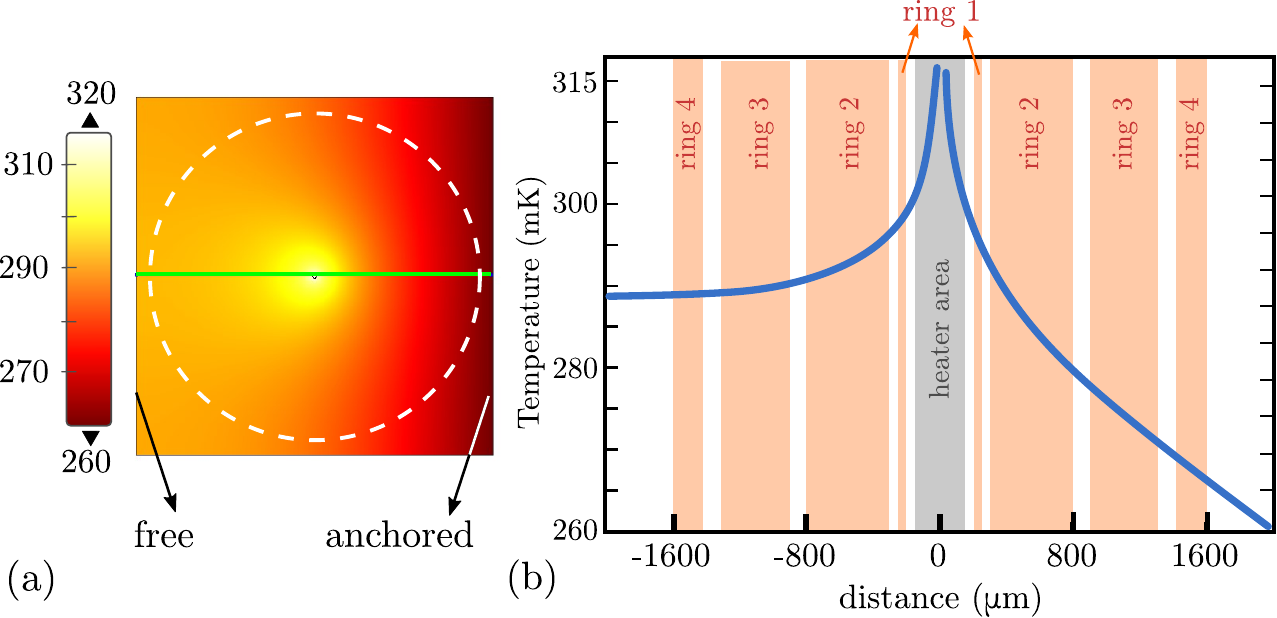}
    \caption{Results of a finite-elements simulation of the temperature distribution. We consider a square sample and a heater power $P = \SI{277}{\nano\watt}$. Given the device construction, only the right edge is anchored to the cold finger at a base temperature of $T = \SI{260}{\milli\kelvin}$. (a)  Colormap of the substrate temperature, the circle in white dotted line indicates the mesa region. (b) Temperature profile along the linear green line indicated in (a). We indicate in gray the heater region and orange corresponds to the different Corbino rings positions.}        
    \label{fig:finiteElem}
\end{figure}

\begin{figure}
    \centering
    \includegraphics[width=\columnwidth]{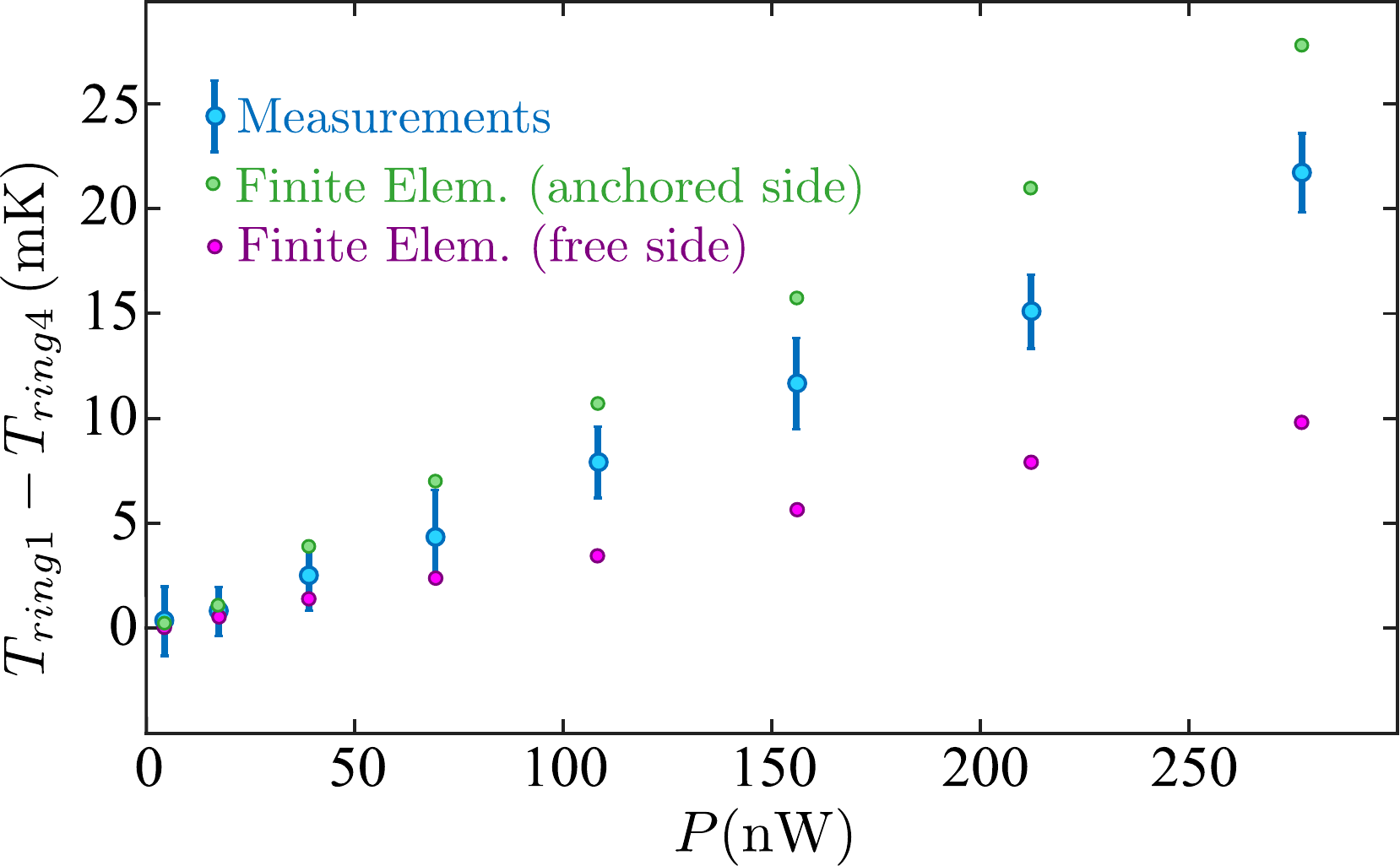}
    \caption{Difference between the  mean local temperatures of the rings $1$ and $4$, as a function of the power applied to the heater. They are obtained from the data of Fig. \ref{fig:calT}, following the procedure described in the text. We also include the results of the finite-elements model, corresponding to Fig.~\ref{fig:finiteElem} for both the right thermally anchored side and the left free side of the sample.} 
    \label{fig:deltaTpower}
\end{figure}

To benchmark the previous estimates for the temperature difference induced in the 2DES we simulate the temperature profile using values of the thermal conductivity $\kappa$ of the substrate previously reported in the literature at \SI{300}{\milli\kelvin}, $\kappa =$ \SI{1.6e-2}{\watt\per\kelvin\per\meter} \cite{chickering2010thermopower, chickering2013thermoelectric} (following a law $\propto T^{2.56}$).
In these works, it was shown that the substrate phonons determine the temperature gradient at mK temperatures as we are assuming here. This is supported by the fact that the estimate for the conductivity of the 2DES is $\SI{4e-21}{\watt\per\kelvin\per\meter}$, i.e., orders of magnitude smaller.

We performed finite-element simulations of the heat-flow equation, its results for a heater power of \SI{277}{nW} are shown in the color-map of Fig.~\ref{fig:finiteElem}. Following the experimental setup, only the  side indicated as anchored is assumed to be thermalized at the temperature of the cold finger, while the other edges have free contour conditions.

The calculated temperature differences are shown in Fig.~\ref{fig:deltaTpower}, together with the experimental results. 
Estimates for the anchored side at a power of \SI{277}{nW}, for instance, is $(T_{\text{ring 1}}-T_{\text{ring 4}})^{\rm sim} = (\Delta T)^{\rm sim} \approx \SI{28}{\milli\kelvin}$, while in the free side, the temperature difference results to be $(\Delta T)^{\rm sim} \approx \SI{10}{\milli\kelvin}$. Averaging these results, as the experiment does when integrating over the ring ohmic contacts, we obtain $(\Delta T)^{\rm sim} = \SI{19}{\milli\kelvin}$, in good agreement with the measured value of $(\Delta T)^{\rm meas} = \SI{22(2)}{\milli\kelvin}$.

To finalize, we also compare the estimates for this temperature difference to the ones inferred from thermoelectric measurements. 
In the latter case, the thermal gradient is generated by applying an ac voltage to the heater with a  period of \SI{36}{ms}, which is much longer than the typical relaxation rate of the electrons.
At this frequency range, the thermal distribution can be modeled by the stationary behavior. 
Deviations are expected only at much higher frequencies.
The resulting current $I(t)$ of amplitude $I_0$ and angular frequency $\Omega$ results in a power given by
$ P(t)= R I_0^2\left[1- \cos(2 \Omega t + 2 \varphi) \right]/2$, being $\varphi$ 
a possible arbitrary phase and $R$ is the resistance of the heater. 
The generated temperature bias between the heater and the external rim of the structure can be assumed to have the following behavior
\begin{equation}\label{eqs3}
\Delta T (t) = \left( \Delta T \right)_0 - \left( \Delta T \right)_2 \cos (2 \Omega t + 2 \varphi),
\end{equation}
with $\left( \Delta T \right)_2 \simeq \left( \Delta T \right)_0$, independent of the frequency.

According to  the measurements of thermal voltage reported in Ref. \onlinecite{real2020thermoelectricity} under similar experimental conditions as in the present work (same device, power of \SI{277}{\nano\watt} and a bath temperature of \SI{267}{\milli\kelvin}), the estimate for the  oscillating component of the temperature bias  $(\Delta T)_2$ for the ring 2 is $(\Delta T)_2 \simeq (\Delta T)_0= \SI{1.78}{\milli\kelvin}$. 
These values are compatible albeit smaller than the  thermal drop calculated with simulations of the flow equation discussed in Fig. \ref{fig:finiteElem}.
The  numerical estimates for the $ring \; 2$ based on these simulations are $(\Delta T)^{\rm sim}  \approx \SI{5.8}{\milli\kelvin} \; (\SI{12.7}{\milli\kelvin})$ for the free (anchored) side, respectively, if we assume the same phononic thermal conductivity as in the results of Fig. \ref{fig:finiteElem}. Instead, calculations with a different value of this quantity
($\kappa =$ \SI{5 e-2}{\watt\per\kelvin\per\meter} \cite{ChickeringPhD}) lead to a temperature difference close to 
the thermoelectric estimate. Deviations by nearly an order of magnitude of  the phonon-thermal-conductivity values caused not only from impurities but also from the effect of the dimensions and the surface-reflectivity of the phonons are not uncommon. In this sense, the presence of intermediate ohmic contacts could also induce deviations. 
This aspect deserves further investigation.

To summarize, we have presented a reliable method for generating a thermal bias in a 2D electron system of a Corbino  configuration in the quantum Hall regime, based on a central heater. 
We have also shown that the measurements of the electrical conductance in thin rings along  the Corbino disk provide 
a convenient mechanism to implement the thermometry to sense the temperature drop. The experimental results are supported by numerical simulations of the heat flux through the substrate. Our analysis was performed in GaAs samples, but the present methodology can be also adapted to study thermal and thermoelectric phenomena in graphene samples. 
This is particularly relevant in light of the active research activity using the Corbino geometry in graphene \cite{yan2010charge,zhao2012magnetoresistance,faugeras2010thermal,zeng2019high,kamada2021strong}. Another avenue of research is the calibration of the measurements of the conductance and the thermovoltage against fixed points or absolute thermometers to fabricate accurate temperature sensors for metrological applications.

%%%%%%%%%%%%%%%%%%%%%%%%%%%%%%%%%%%%%%%%%%%%%%%%%%%%%%%%%%%%%%%%%%%%%%%%%%%%%%%%
% \section*{Acknowledgment}
%%%%%%%%%%%%%%%%%%%%%%%%%%%%%%%%%%%%%%%%%%%%%%%%%%%%%%%%%%%%%%%%%%%%%%%%%%%%%%%%
\begin{acknowledgments}
We thank Klaus von Klitzing for his constant interest and support. J\"urgen Weis (MPI, Germany) for suggestions and help, also from MPI Achim G\"uth and Marion Hagel for the wafer lithography. We thank Lars Tiemann for providing the measurement software. 
We acknowledge support from CONICET, CNEA and INTI, Argentina, MincyT under PICT-2017-2726, PICT-2018-04536 and PICT-2020-03661 from Argentina (LA) and the Swiss National Foundation (Schweizerischer Nationalfonds) NCCR "Quantum Science and Technology" (WD, WW).  
\end{acknowledgments}

\section{Author Declarations}
The authors have no conflicts to disclose.

The data that support the findings of this study are available from the corresponding author upon reasonable request.

\section{Author Contributions}
M.A. Real: 
    Conceptualization; 
    Data curation;
    Formal analysis;
    Investigation;
    Methodology; 
    Software; 
    Writing – original draft; 
    Writing – review \& editing.
A. Tonina: 
    Conceptualization; 
    Funding acquisition;
    Project administration;
    Resources;
    Supervision;   
    Writing – review \& editing.    
L. Arrachea:
    Conceptualization; 
    Formal analysis;
    Methodology; 
    Supervision;
    Writing – review \& editing.
P. Giudici:
    Writing – original draft; 
    Writing – review \& editing.
C. Reichl:
    Resources;
    Writing – review \& editing.
W. Wegscheider:
    Resources;
    Funding acquisition;
    Writing – review \& editing.
W. Dietsche:
    Conceptualization; 
    Investigation;
    Resources;
    Supervision;
    Writing – review \& editing.

\section{References}
\bibliography{Corbino_Temp_meas}% Produces the bibliography via BibTeX.

\end{document}